\documentclass[11pt,a4paper,english]{article}
\usepackage[]{fontenc}
\usepackage[latin1]{inputenc}
\usepackage{graphicx}
\usepackage{setspace}
\makeatletter

 \newcommand{\lyxaddress}[1]{
   \par {\raggedright #1 
   \vspace{1.4em}
   \noindent\par}
 }
 \newenvironment{lyxlist}[1]
   {\begin{list}{}
     {\settowidth{\labelwidth}{#1}
      \setlength{\leftmargin}{\labelwidth}
      \addtolength{\leftmargin}{\labelsep}
      }}
   {\end{list}}

\usepackage{babel}
\usepackage{setspace}

\usepackage{babel}
\makeatother
\begin{document}

\title{GHz sandwich strip inductors based on FeN films}

\author{A. Gromov, V. Korenivski, and D. Haviland}

\maketitle

\lyxaddress{\centering \emph{Nanostructure Physics, Royal Institute of Technology,
10044 Stockholm, Sweden.}}

\begin{abstract}
Planar strip inductors consisting of two \emph{Fe-N} films enclosing
a conducting film made of \emph{Cu}, were fabricated on oxidized \emph{Si}
substrates. The inductors were $1\, mm$ long, 2 to 100 $\mu m$ wide,
with layers of thickness $\sim0.1\,\mu m$ for the magnetic films
and $\sim0.5\,\mu m$ for the conductor. The soft ($H_{c}=4-8\, Oe$)
magnetic layers were biased during impedance measurement by applying
an external field along the strip length thereby facilitating the
transverse susceptibility configuration. Biased strips exhibited 70
to 100\% inductance enhancement at $1\, GHz$ with quality factors
$Q$=4.5 to 3, respectively. The magnetic contribution to the total
flux in the narrow devices was less than predicted theoretically,
which was attributed to hardening of the magnetic material at the
edges of the strip, where the deposition was close to 60$°$ incidence.
Test films were fabricated on tilted substrates and found to develop
a very high anisotropy (up to $1\, kOe$) for deposition angles larger
than 30$°$. Optimizing the flux closure at the strip edges and using
thicker conductor layers is essential for further improving the performance
of sandwich strip inductors.\newpage
\end{abstract}

\section{Introduction}

There have been numerous efforts to enhance the performance and diminish
the size of planar inductors for sub-GHz applications by using magnetic
films. Several recent reports extended these efforts to 1-2 GHz by
studying magnetically enhanced spirals and strip inductors. A magnetically
clad spiral showed 10-20\% inductance enhancement \cite{yam1}. An
inductance gain of $\sim$70\% was observed at 1 GHz for a magnetic
sandwich strip, though with $Q\sim1$ \cite{saito}. The inductor
is a fundamental electronic component, which is least compatible with
silicon integration. Achieving high specific inductance values is
crucial for miniaturization of inductors used in filters, oscillators,
matching networks, etc., in high frequency integrated circuits. The
status of research in the field has recently been reviewed \cite{review}.
A GHz magnetic inductor having a substantial inductance gain (preferably
more than a factor of 2) over the corresponding air-core design and
acceptable efficiency ($Q>3$) at \textasciitilde{}1 GHz has not yet
been demonstrated. We report on the fabrication of magnetic/conductor
sandwich strip inductors using Fe-N films. The inductors exhibit a
2-fold inductance gain with $Q\approx3$ at 1 GHz, a substantially
better performance than reported previously. We analyze the factors
limiting the magnetic efficiency and high frequency performance of
these devices.

\section{Sample preparation and measurements}

The inductors are strips of length 1.15 mm with width varying from
2 to 100 $\mu m$. The inductors were fabricated on oxidized Si substrates
(1 $\mu m$ thick oxide layer) with a lift off process. A two layer
resist system consisting of a low contrast bottom layer (PMGI) and
a high contrast top layer (ZEP) was used. The pattern was defined
using an e-beam writer. The mask was developed to have an undercut
of 0.5-1.5 $\mu m$, as shown schematically in Fig. 1. The following
deposition sequence was used: sputtering 0.1 $\mu m$ of \emph{FeN},
evaporating 0.5 $\mu m$ of \emph{Cu}, sputtering 0.1 $\mu m$ of
\emph{FeN}. A 5 nm thick capping layer of gold was evaporated to ensure
a good contact for impedance measurements. Magnetic material covering
the edges of the strips (flanges) was achieved by utilizing the difference
in shadow depth for sputtering of the magnetic layers (dashed arrows)
and evaporation of the conductor (solid arrows). A magnetic/conductor/magnetic
sandwich with magnetic flux closure at the edges is thus obtain with
\emph{one} lift-off mask.

The Fe-N was produced by reactive sputtering from an Fe target using
an argon/nitrogen gas mixture with typically 4\% partial pressure
of nitrogen ($p_{N}$). The sputtering pressure used was 3 \emph{mTorr}
and the rate was 4.8 \emph{Å/s}. Thus sputtered, the Fe-N material
had a resistivity of 40 $\mu\Omega\cdot cm$ as measured using a four-point
technique. Vibrating sample magnetometer (VSM) measurements show a
square loop with coercivity $H_{c}$=4-8 \emph{Oe,} saturation magnetization
$4\pi M_{s}=17\, kGs$, and essentially no in-plane anisotropy. The
copper film was evaporated at 10 \emph{Å/s} and showed the resistivity
of 2.4 $\mu\Omega\cdot cm$. The depositions were made at ambient
temperature.

An HP8714C network analyzer with a Cascade 12 GHz fixed-pitch probe
was used for impedance measurements. The real and imaginary parts
of the impedance were obtained from reflection measurements in the
30 MHz-3 GHz frequency range. The following de-embedding procedure
was used. The network analyzer was calibrated to the probe plane.
A correction was made for the electrical delay corresponding to the
probe length. The contact resistance was evaluated from the low frequency
limit for various device geometries. The resistance scaled linearly
versus the inverse strip width with a $0.24\,\Omega$ offset, which
was attributed to contact resistance. The Cascade probe introduces
a series impedance, which was determined using Cu test strips of varying
width. The impedance of these strips could be accurately modeled and
compared with measurements. The parallel impedance due to the measurement
configuration was found to be negligible for strip widths greater
that 2 $\mu m$.

\section{Results and discussion}

The inductance, $L=Im(Z)/\omega$, is plotted in Fig. 2 as a function
of frequency for different inductor widths. The magnetic films have
no preferred in-plane orientation of the magnetization. The weak,
if any, in-plane anisotropy can result in an arbitrary in-plane orientation
of the magnetic moments (domains) in the films when no biasing field
is applied. This would diminish the magnetic response and limit the
frequency range of these devices, as observed (dashed lines, Fig.
2). To avoid these undesirable effects, we biased the films in an
external magnetic field parallel to the strip length. In the biased
case, we observe a flat response out to the ferromagnetic resonance
(FMR) cutoff (Fig. 2, solid lines). In what follows, we will discuss
only the biased response.

\subsection{Magnetic reluctance analysis}

From the low frequency limit of the inductance (Fig. 2, solid lines)
we can determine the magnetic contribution, which excludes any dissipation
or resonance effects at high frequencies. The total inductance in
this limit, $L=L_{0}+L_{m}$, is the sum of the air-core inductance
of the strip, $L_{0}$, and the contribution due to the magnetic film,
$L_{m}$. In Fig. 4, $L_{m}$ is plotted versus the strip width. The
experimental data deviate from the ideal $1/width$ (dash-dotted line)
behavior expected when there is perfect magnetic flux closure at the
edges. The observed magnetic contribution to the inductance is below
the theoretical value for narrow strips, while it approaches the expected
values for wide strips. This behavior indicates an incomplete flux
closure at the edges, which plays a progressively larger role as the
width of the strip is reduced \cite{vk}.

In order to gain insight into the magnetic state of the flange we
model the data in Fig. 3 using the theory of Ref. \cite{vk} for three
different cross sectional geometries: a flanged edge, no-flange, and
a three section arbitrary edge-gap. In the case of the open structure
(no flanges, dashed line, Fig. 3) the relative permeability is the
only fitting parameter. The fit is poor, indicating that some flux
closure at the edges does take place. Fig. 4a is a SEM image of the
inductor edge showing that the magnetic film becomes thinner and possibly
is discontinuous towards the very edge sloped at approximately 60$°$.
This edge geometry is indeed an intermediate case between perfect
flux closure and no flux closure. To model this case we use a model
geometry depicted in Fig. 4b \cite{vk}. Assuming the flange width
is the same as the thickness of the magnetic films and assuming the
films and flange have the same permeability, we fit the width dependence
of the inductance by introducing a flange gap. A good fit of the experimental
data is obtained for $\mu_{r}=390$ and the flange gap of $g=44nm$.
It should be noted that the effect on the inductance of introducing
a gap in the flange is similar to the effect of a reduced permeability
of a continuous flange. In any case, our data clearly exhibit an incomplete
flux closure at the edges.

\subsection{Magnetic properties of the flange}

Because the cross section of the sandwich does not appear to have
any gap, we suspect that poor flux closure is actually due to lower
permeability of the flange. To investigate this effect, we varied
the angle of the substrate with respect to the source during the evaporation
of the conductor through the lift-off mask. In this way we can vary
the slope of the conductor edge between approximately $\theta=$45$°$
and 80$°$. Fig. 4a is a SEM image of the inductor cross-section showing
that the magnetic film towards the very edge is sloped at approximately
60$°$ to the strip's plane and becomes slightly thinner. For the
top film deposition the incident flux of Fe-N is perpendicular to
the \emph{Cu} surface accept at the edges, where the angle of incidence
is in this case 60$°$. Deposition of soft magnetic materials, including
\emph{Fe-N}, on tilted substrates is known to result in a strong magnetic
anisotropy perpendicular to the direction of incidence \cite{ozawa}-\cite{tim}.
In addition, films deposited at large oblique angles are typically
thinner. In order to quantify the magnetic state at the edge, we have
fabricated test \emph{Fe-N} films sputtered directly on oxidized Si
at 0$°$ to 60$°$ angle of incidence. The results of \emph{VSM} measurements
for these films are shown in Fig. 5. The normal incidence films show
no preferred in-plane orientation of the magnetization. The films
deposited at 20$°$ incidence develop a weak in-plane anisotropy.
The easy-axis coercivity, $H_{ce}$, increases while that for the
hard axis, $H_{ch}$, decreases, and a uniaxial anisotropy with $H_{k}\approx10\, Oe$
appears. The films sputtered at $\theta=$ 60$°$ (inset) show $H_{ce}\approx200\, Oe$
and $H_{k}\approx400\, Oe$. For the films sputtered at $p_{N}=4-8\%$,
$H_{ce}$ is the same, while anisotropy increases up to $H_{k}=600-800\, Oe$.
Increasing the $\theta$ to the 80$°$, we obtain for $H_{k}=0.9-1\, kOe$.
This growth induced anisotropy is directed along the length of the
strip, leading to more than an order of magnitude reduction in the
transverse permeability, and thus degrading the inductive response
of the film. Two possible mechanisms for this anisotropy are discussed
in the literature (see, e.g., \cite{tim}). The first is the shape
anisotropy of elliptical magnetic columns that are typically present
in films deposited at oblique angles due to the self-shadowing effect
\cite{jo,ozawa}. This effect has recently been analyzed in detail
for Co films on underlayers deposited at oblique angles \cite{McMichael}
and was shown to result in up to 1 kOe of anisotropy. Another possible
origin of the oblique-growth induced anisotropy in magnetic films
discussed in the literature is due to magnetostriction. Although we
have not performed independent measurements of the stress in our films
as a function of the incidence angle, the amount of stress needed
to produce such high anisotropy values appears unreasonable if one
takes typical values for the magnetostriction of Fe-N.

In order to directly verify the effect of the flange on the inductance
we have fabricated two inductor sets under identical conditions, varying
only the \emph{Cu} edge angle, $\theta=$45$°$ and 80$°$. The magnetic
contribution to the total inductance is shown in Fig. 6. The difference
in $L_{m}$ for the two inductor sets increases for small widths,
approaching 25\% for the $5\,\mu m$ wide strip. We attribute this
increase in $L_{m}$ to a magnetically softer flange in the 45$°$-edge
inductor, resulting in a better flux closure.

\subsection{HF performance}

The contribution of the flanges to the magnetic inductance becomes
smaller as the width of the strip is increased (see Fig. 3). The measured
inductance of the $50\,\mu m$ wide strip is essentially the value
expected for a sandwich with perfect flux closure, hence the model
for the impedance of a closed magnetic structure should apply (see
Appendix and \cite{grom}). The inductance and quality factor for
the $50\,\mu m$ width strip measured in three biasing fields are
shown as functions of frequency in Fig. 7a and 7b, respectively, along
with the theoretically fitted curves. Fixing the geometrical parameters
at the measured values, a good fit can be obtained by varying the
permeability and intrinsic damping. If we use the same parameters
to fit the impedance of the $10\,\mu m$ strip, the deviation from
the predicted performance is significant (see Fig. 8). In order to
obtain a good fit not only the permeability has to be significantly
reduced (by roughly a factor of two) but the dissipation constant
has to be significantly increased: from $\kappa=$0.028 (consistent
with values 0.020-0.025 measured on test films) to 0.06. The reduction
in the effective permeability for narrow strips can be attributed
to the increasing influence of incomplete flux closures at the edges.
However, the increase in the damping constant is not expected to depend
on the width of the strip in the simplest picture. A probable cause
of the increased dissipation is that an increasing portion of the
magnetic flux leaks through the conductor near the flange, causing
screening currents in the conductor. This dissipation mechanism is
not accounted for in the 'closed-flux' model \cite{grom} but known
to be the dominating loss mechanism in sandwiches without flux closure
\cite{sukstan}.

The inductance and quality factor at 1 GHz are shown in Fig. 9 as
a function of the inductor width. A 2-fold inductance gain with $Q\approx3$
was obtained for the $20\,\mu m$ wide inductor. By varying the width
of the device the inductance gain can be compromised for higher quality
factor.

\section{Conclusions}

We have demonstrated a significant specific inductance gain at 1 GHz
by using magnetic/conductor sandwiches. Several issues should be addressed
for further improving the device performance. The use of soft films
with resistivities in access of 100 $\mu\Omega$-cm should allow thicker
magnetic layers in the sandwich, resulting in larger magnetic flux
and therefore larger inductance, without increasing dissipation. Use
of thicker conductor layers should yield higher quality factors, provided
the edge flux closures are not thereby degraded. Control of the properties
of the magnetic films at the strip edges to achieve efficient flux
closure is crucial for both improving the inductance gain and quality
factor in the GHz range. Inducing a longitudinal in-plane anisotropy
in the films, for example by field annealing, should eliminate the
need for external biasing.

\section{Appendix}

Here we reduce the full solution for the impedance of a 'closed' sandwich
inductor \cite{grom} to a simple expression, which should be useful
in analyzing or designing sandwich inductors without insulation layers.
Regrouping the impedance from \cite{grom} in such a way as to separate
the contribution of the conductor one obtains:\[
Z\left(\omega\right)=\frac{\ell}{w\left(b\sigma_{0}+2d\sigma_{1}\right)}+i\omega\frac{\mu_{0}\ell}{2\pi}\left(\ln\left(\frac{2\ell}{w}\right)+1\right)+i\omega\,\frac{\mu_{1}\ell\,\delta_{1}}{2w\left(1+i\right)}\tanh\left(\left(1+i\right)\frac{d}{\delta_{1}}\right)\, k\!\left(\omega\right)\]
 where $\delta_{0(1)}=\sqrt{\frac{2}{\omega\mu_{0(1)}\sigma_{0(1)}}}$
is the skin depth in region {}``0'' ({}``1''). The first term
here is due to the parallel connection of the conductor and magnetic
film resistances, $R_{0}=\frac{R_{c}R_{m}}{R_{c}+R_{m}}$. The second
term is the air-core reactance $i\omega L_{0}$. The last term represents
the magnetic contribution, including the skin effect. The complex
coefficient, \[
k\!\left(\omega\right)=\frac{\sqrt{\frac{\mu_{1}\sigma_{0}}{\mu_{0}\sigma_{1}}}\tanh\left(\left(1+i\right)\frac{b}{2\delta_{0}}\right)+\coth\left(\left(1+i\right)\frac{d}{\delta_{1}}\right)}{\sqrt{\frac{\mu_{1}\sigma_{0}}{\mu_{0}\sigma_{1}}}\tanh\left(\left(1+i\right)\frac{b}{2\delta_{0}}\right)+\tanh\left(\left(1+i\right)\frac{d}{\delta_{1}}\right)}-\frac{\coth\left(\left(1+i\right)\frac{d}{\delta_{1}}\right)}{\left(1+i\right)\frac{d}{\delta_{1}}\left(1+\frac{b\sigma_{0}}{2d\sigma_{1}}\right)}\]
 is the correction factor arising from the current redistribution
between the conductor and magnetic layers. When $d<b$$,$ $\sigma_{1}\ll\sigma_{0}$,
$k$ becomes very close to unity, and the largest error introduced
by omitting this coefficient is less than $\frac{2d\sigma_{1}}{b\sigma_{0}}$,
which is small at all frequencies. Thus, we can express the impedance
using the equivalent circuit of Fig. 10 as \[
Z\left(\omega\right)=R_{0}+i\omega L_{0}+i\omega\, L_{m}\!\left(\omega\right),\]

where \[
R_{0}=\frac{R_{c}R_{m}}{R_{c}+R_{m}}=\frac{\ell}{w\left(b\sigma_{0}+2d\sigma_{1}\right)},\]

\[
L_{0}=\frac{\mu_{0}\ell}{2\pi}\left(\ln\left(\frac{2\ell}{w}\right)+\frac{1}{2}\right)\]
 is the exact air-core inductance \cite{Ldc} differing by $\sim$10\%
from the approximation used in \cite{grom}, and

\[
L_{m}\!\left(\omega\right)=\frac{\mu_{1}\ell\,\delta_{1}}{2w\left(1+i\right)}\tanh\left(\left(1+i\right)\frac{d}{\delta_{1}}\right)\]
 is the inductance of a magnetic film in a uniform magnetic field
\cite{riet}. The effective slab thickness is $2d$, which at first
glance might appear counter intuitive, but is a natural result of
current redistribution in a sandwich with the layers in direct electrical
contact. The intrinsic transverse permeability $\mu_{1}$ is the well
known Landau-Lifshitz permeability (see, e.g., \cite{usov}):

\begin{equation}
\mu_{1}=\mu_{0}\frac{\left(\omega_{H}+\omega_{M}\right)^{2}-\omega^{2}}{\omega_{H}\left(\omega_{H}+\omega_{M}\right)-\omega^{2}},\label{transv}\end{equation}
\begin{equation}
\omega_{M}=4\pi M_{s}\gamma,\quad\omega_{H}=\gamma\left(H_{0}+H_{K}\right)-i\omega\kappa,\label{MH}\end{equation}
 where $\gamma$ is the gyromagnetic ratio, $\kappa$ - fenomenological
damping parameter, $M_{s}$ - saturation magnetization, $H_{K}$ -
uniaxial anisotropy field in z-direction, and $H_{0}$ - external
uniform magnetic field. 
\newpage

\section*{Figure captions:}

\begin{lyxlist}{00.00.0000}
\item [Fig.1.]Schematic of the fabrication process. 
\item [Fig.2.]Inductance as a function of frequency for 5, 10, 20, 50,
and 100 $\mu m$ wide strips with $H_{bias}=0$ (dashed lines) and
$H_{bias}=35$ Oe (solid lines). 
\item [Fig.3.]The magnetic contribution to inductance, $L_{m}$. The expected
inductance of a 'closed' magnetic structure (dash-dot line) and an
'open' structure (dashed line). 
\item [Fig.4.]Cross-sectional view of the sandwich inductor (a). Modeling
geometry (b): $\mu=\mu_{0}\mu_{r}$, $t_{m}$ - magnetic film thickness,
$w_{f}=t_{m}$ - flange width, $g$ - gap. 
\item [Fig.5.]Hard and easy axis M-H curves for Fe-N films deposited at
different oblique angles. 
\item [Fig.6.]Magnetic inductance versus width for two sets of inductors
having the conductor edge angle of approximately $\theta=$45 and
80 degrees to the film normal. 
\item [Fig.7.]L (a) and Q (b) versus frequency for a 50 um wide inductor
in three biasing fields. Solid lines are theoretical fits using the
'closed' model \cite{grom}. 
\item [Fig.8.]L (a) and Q (b) versus frequency for a 10 um wide inductor
in two biasing fields. Solid lines are theoretical fits using \cite{grom}
with the parameters from Fig. 7. Dashed lines are fits assuming a
reduced permeability and increased damping constant (see text). 
\item [Fig.9.]Inductance and quality factor at 1 GHz versus inductor width. 
\item [Fig.10.]An equivalent circuit of a magnetic sandwich inductor without
insulation layers.
\end{lyxlist}
\begin{figure}
\includegraphics[%
  width=1.0\columnwidth]{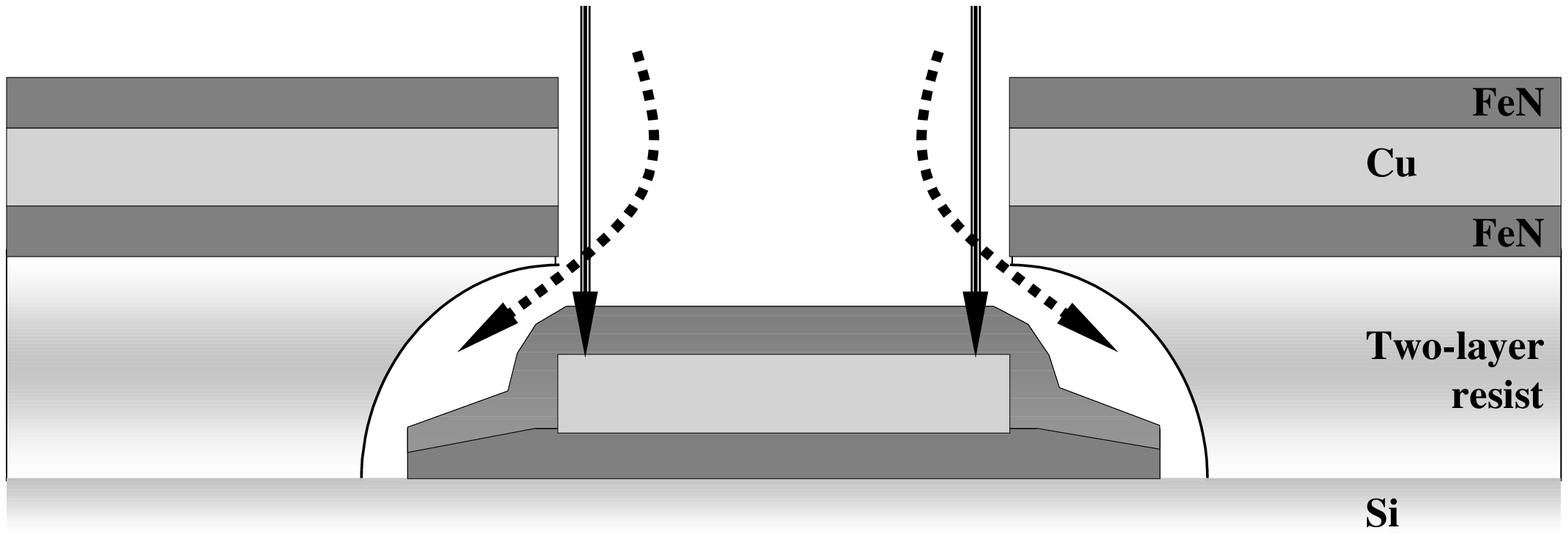}

\caption{-}
\end{figure}

\begin{figure}
\includegraphics[%
  width=1.0\columnwidth]{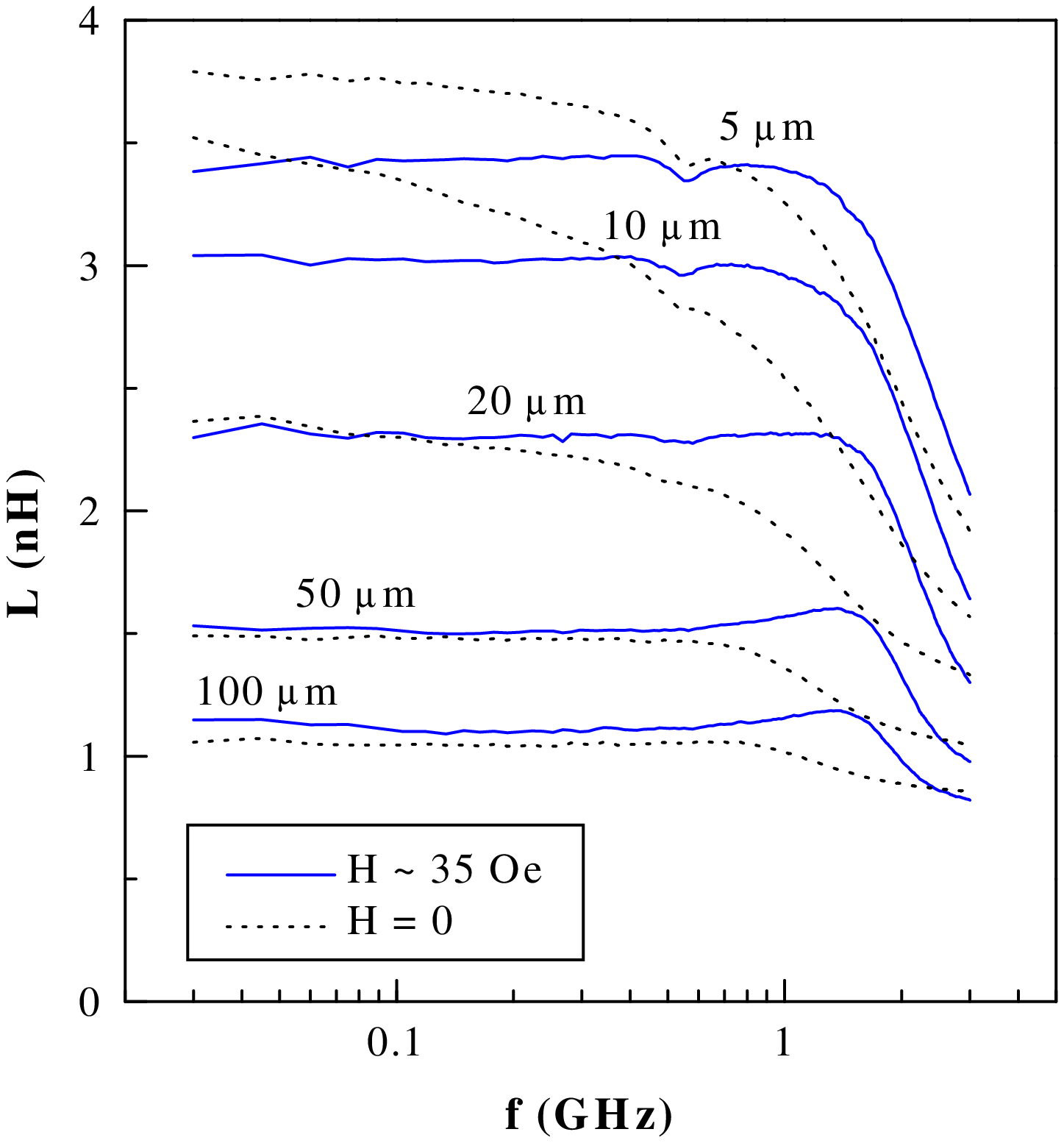}

\caption{-}
\end{figure}

\begin{figure}
\includegraphics[%
  width=1.0\columnwidth]{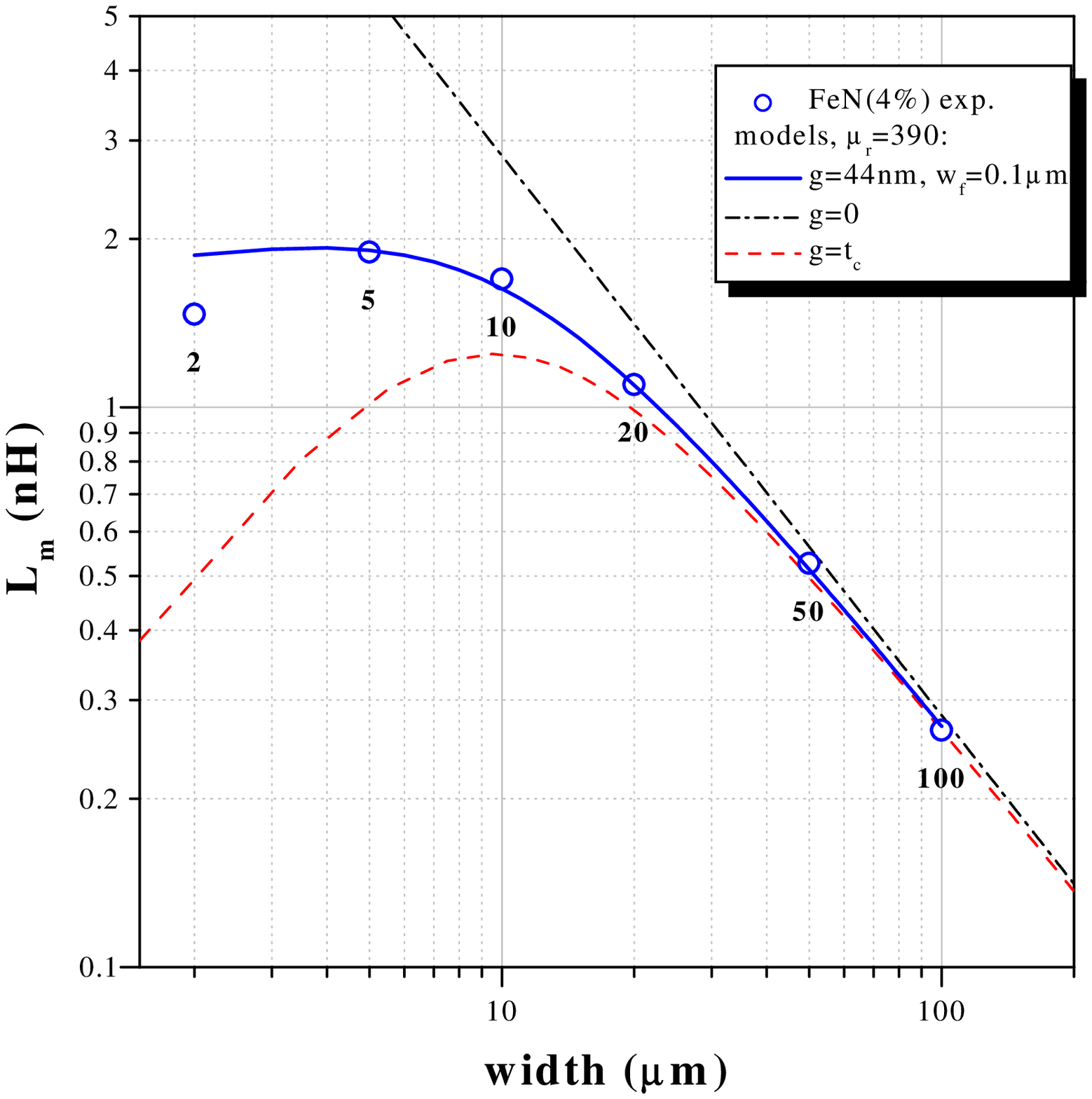}

\caption{-}
\end{figure}

\begin{figure}
\includegraphics[%
  width=1.0\columnwidth]{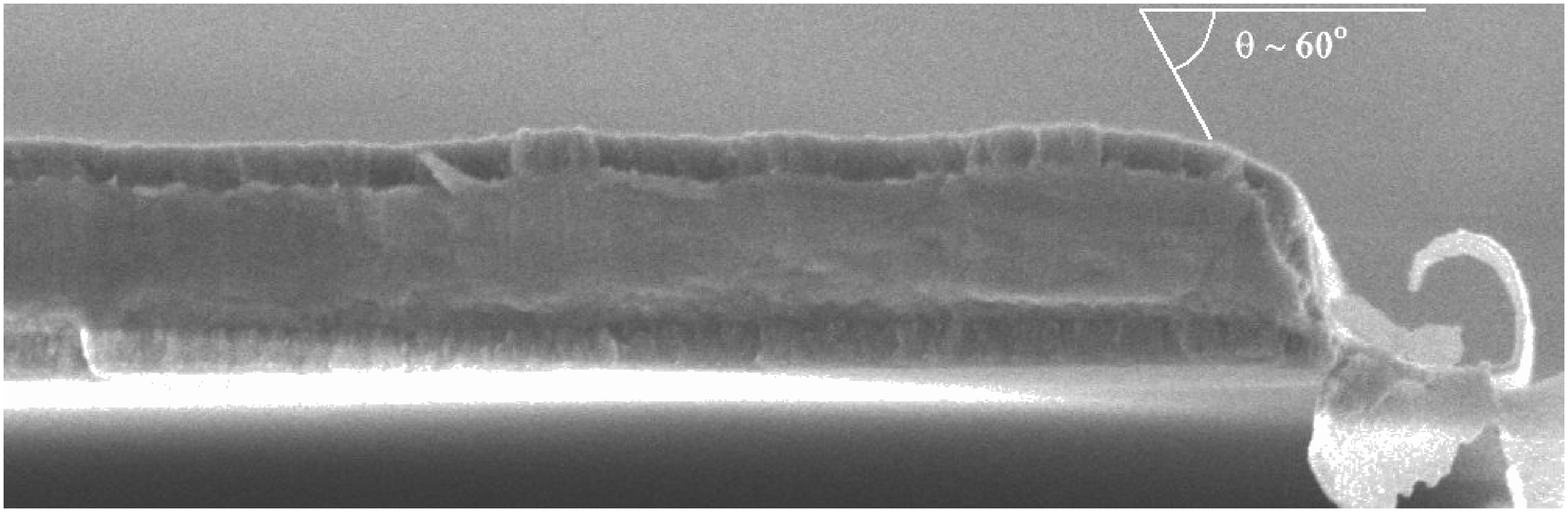}

\includegraphics[%
  width=1.0\columnwidth]{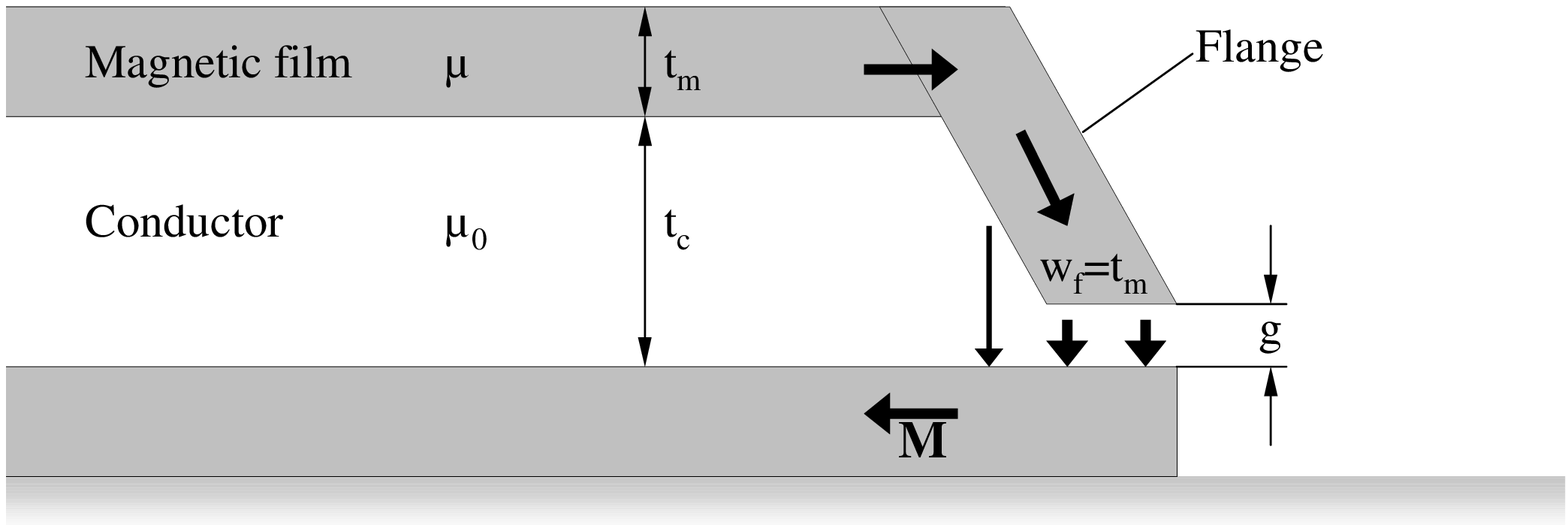}

\caption{-}
\end{figure}

\begin{figure}
\includegraphics[%
  width=1.0\columnwidth]{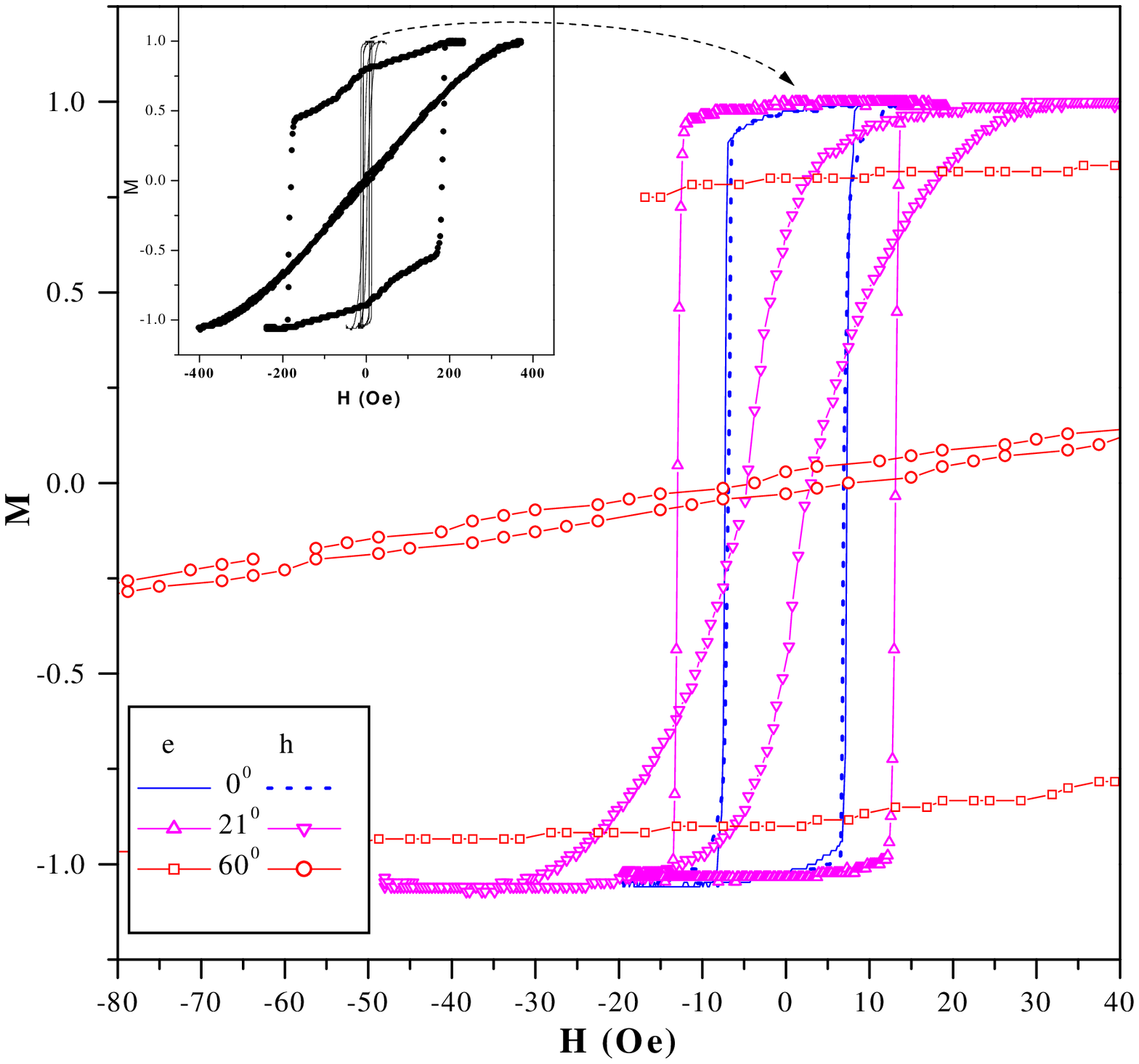}

\caption{-}
\end{figure}

\begin{figure}
\includegraphics[%
  width=1.0\columnwidth]{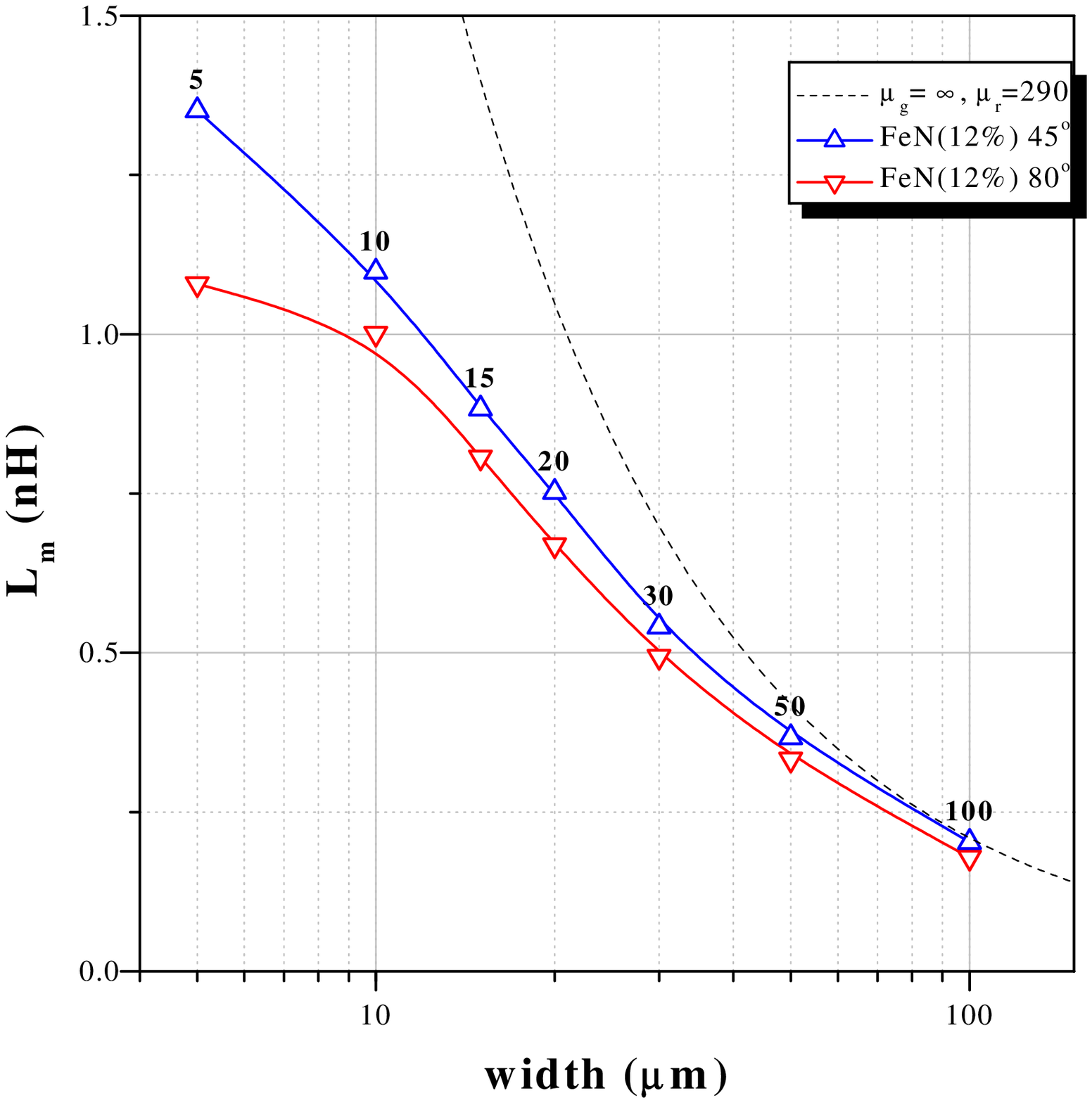}

\caption{-}
\end{figure}

\begin{figure}
\includegraphics[%
  width=0.90\columnwidth]{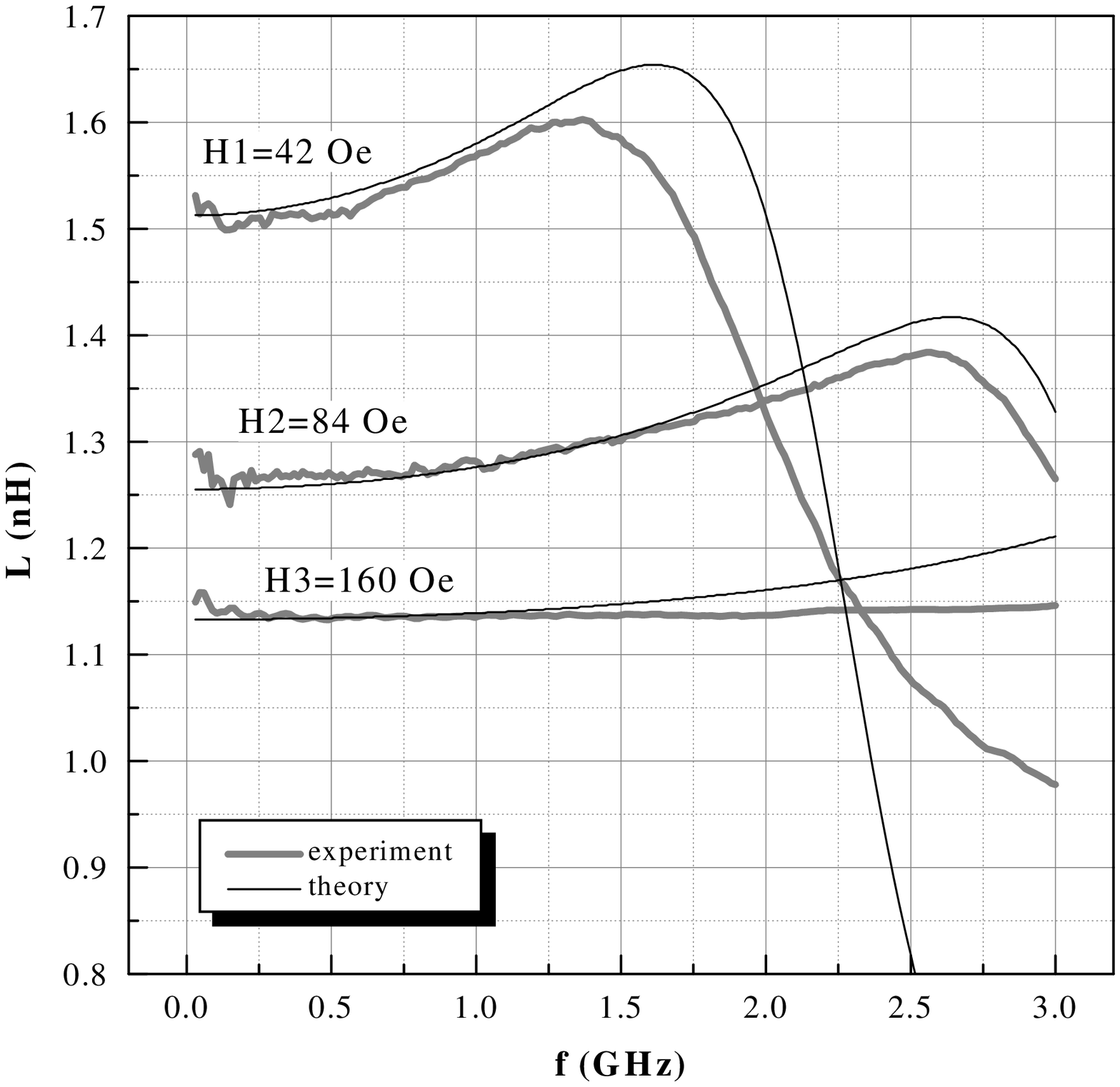}

\includegraphics[%
  width=0.90\columnwidth]{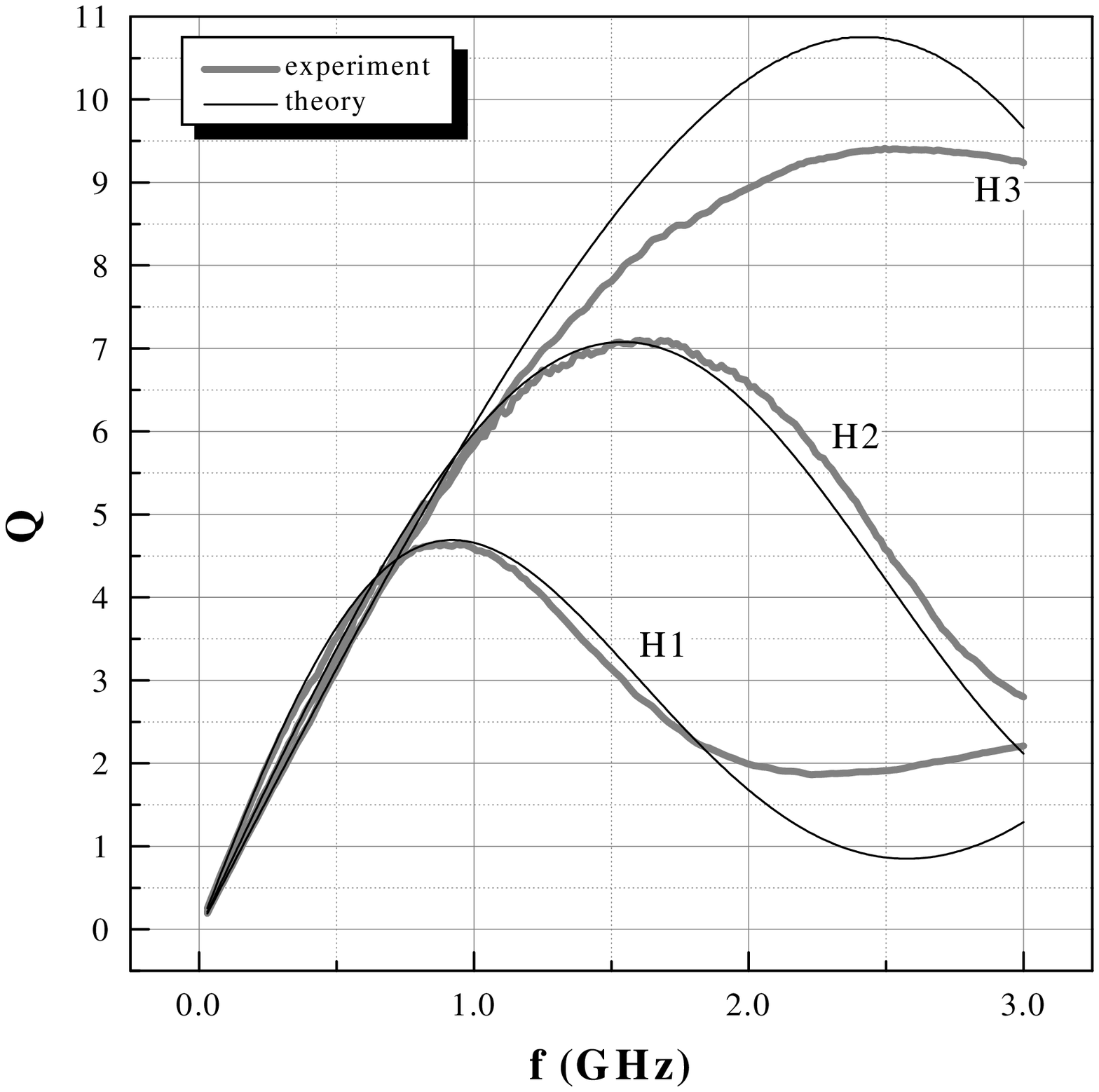}

\caption{-}
\end{figure}

\begin{figure}
\includegraphics[%
  width=0.90\columnwidth]{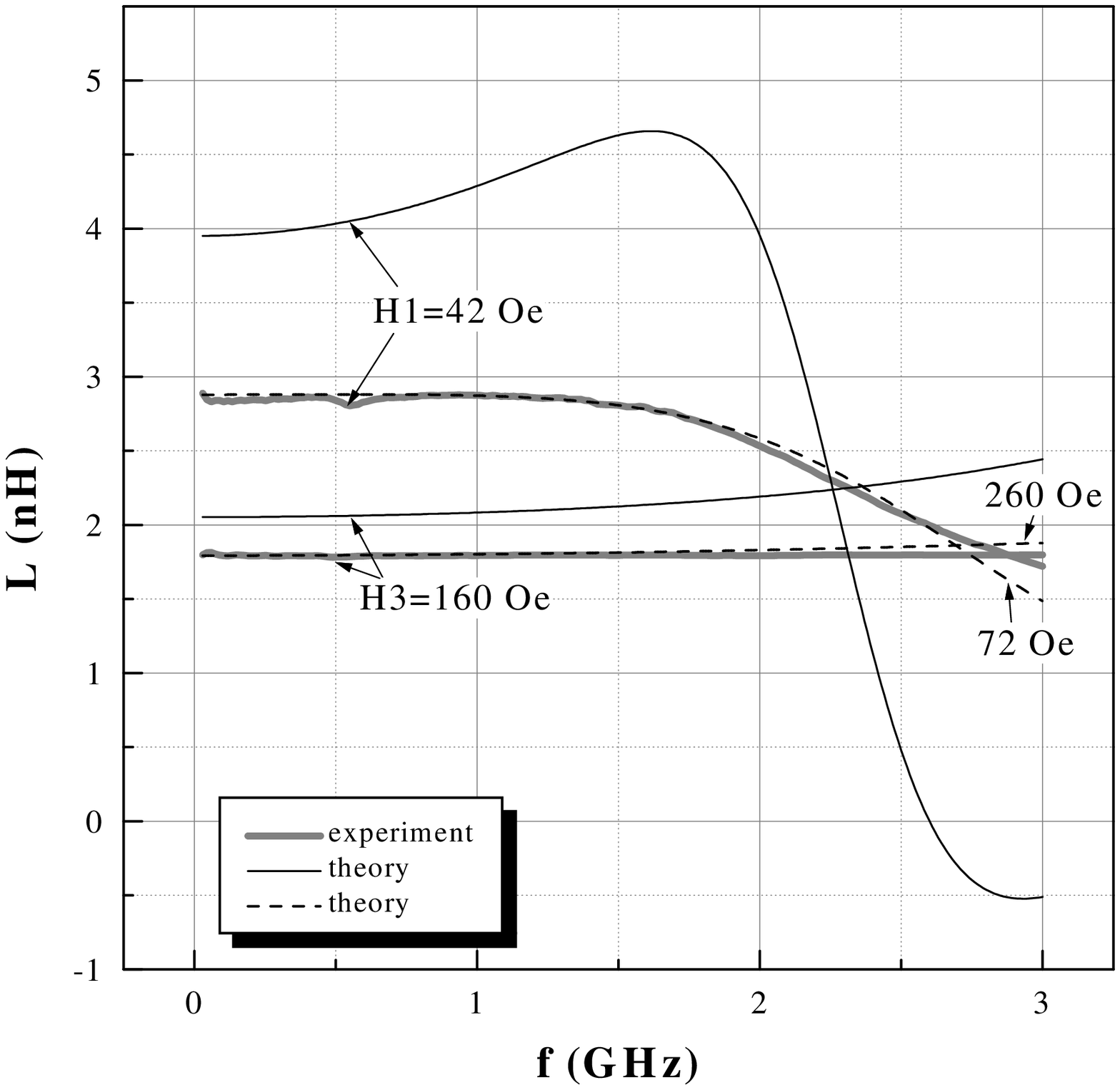}

\includegraphics[%
  width=0.90\columnwidth]{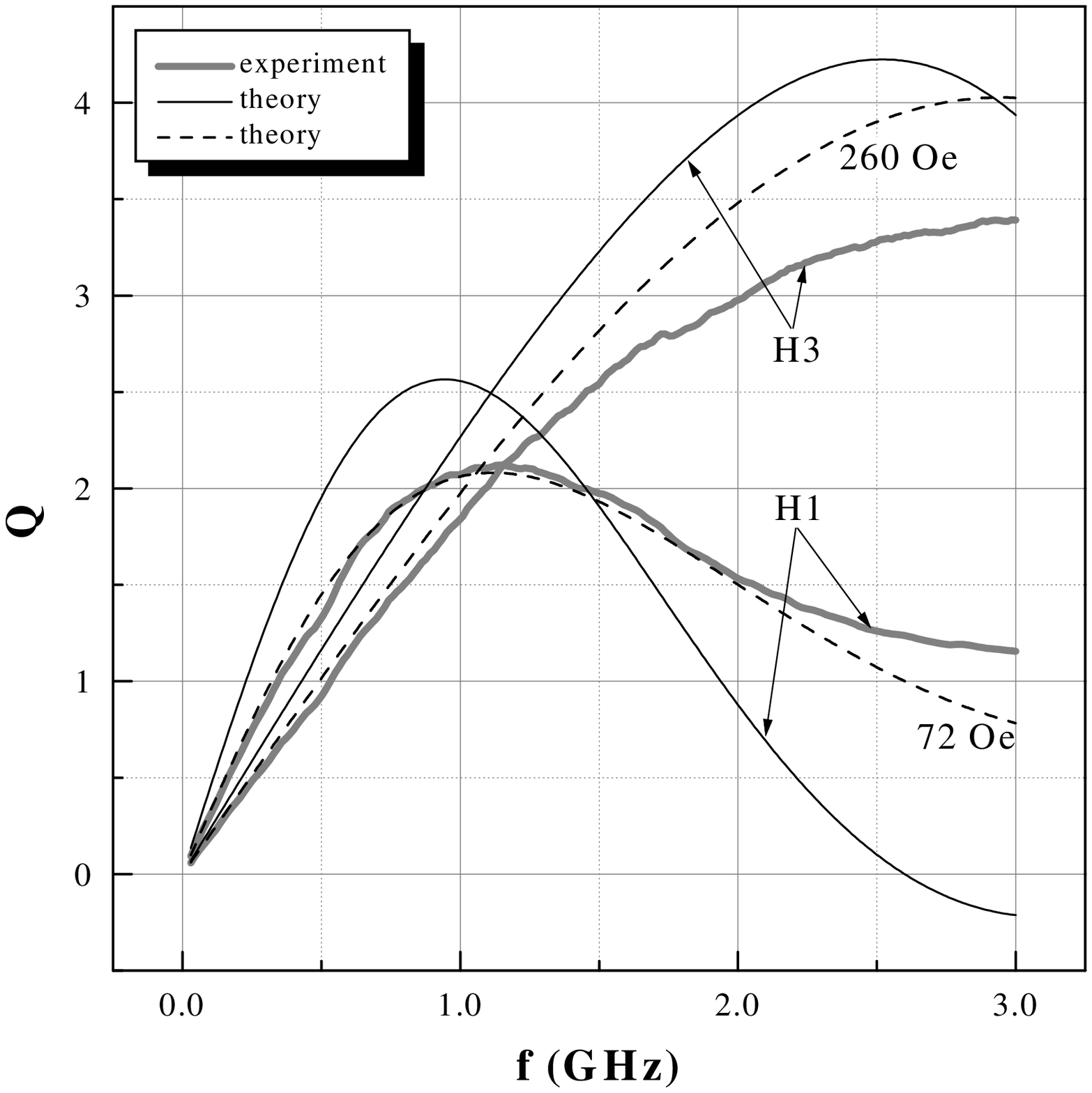}

\caption{-}
\end{figure}

\begin{figure}
\includegraphics[%
  width=1.0\columnwidth]{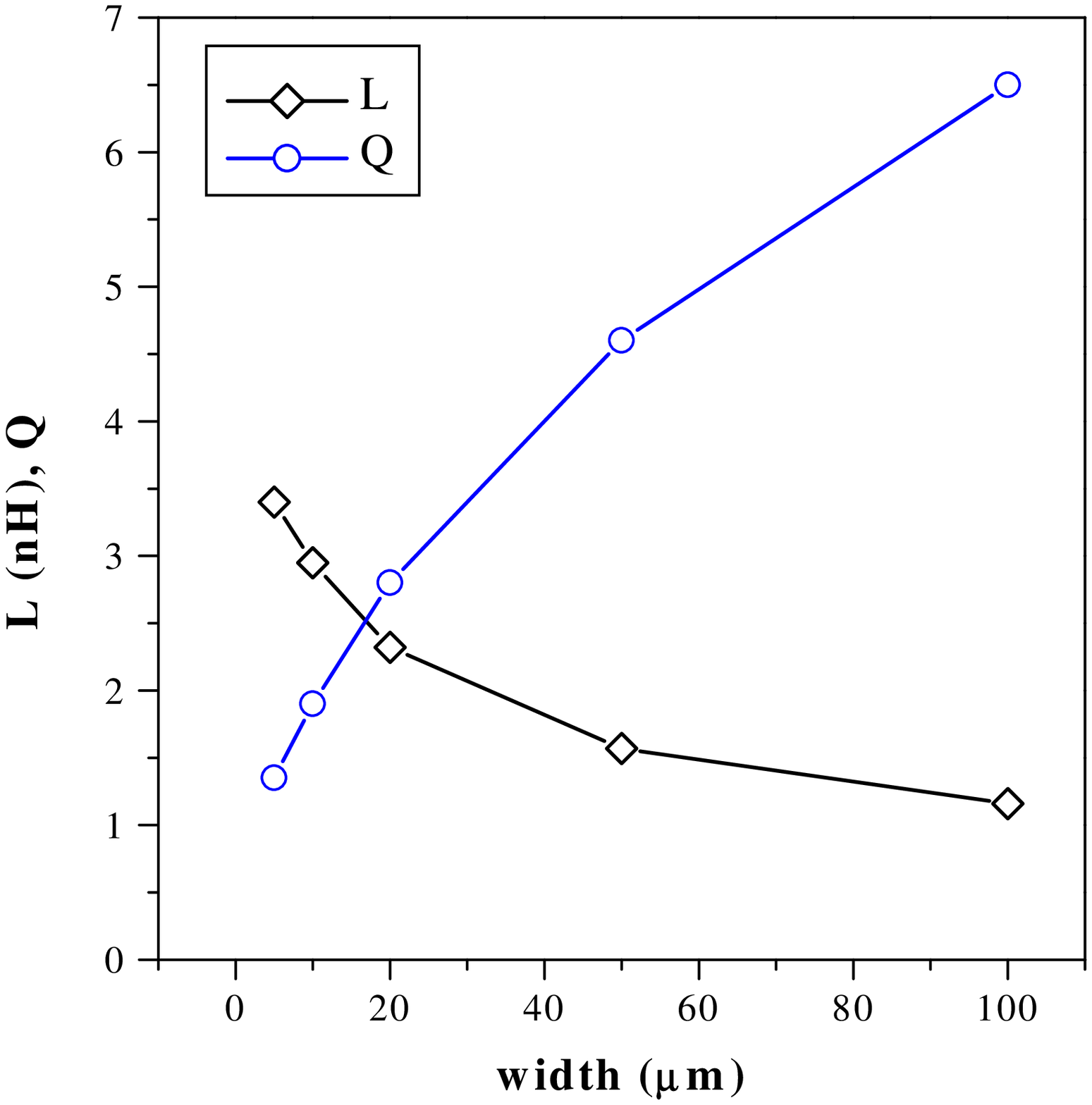}

\caption{-}
\end{figure}

\begin{figure}
\includegraphics[%
  width=1.0\columnwidth]{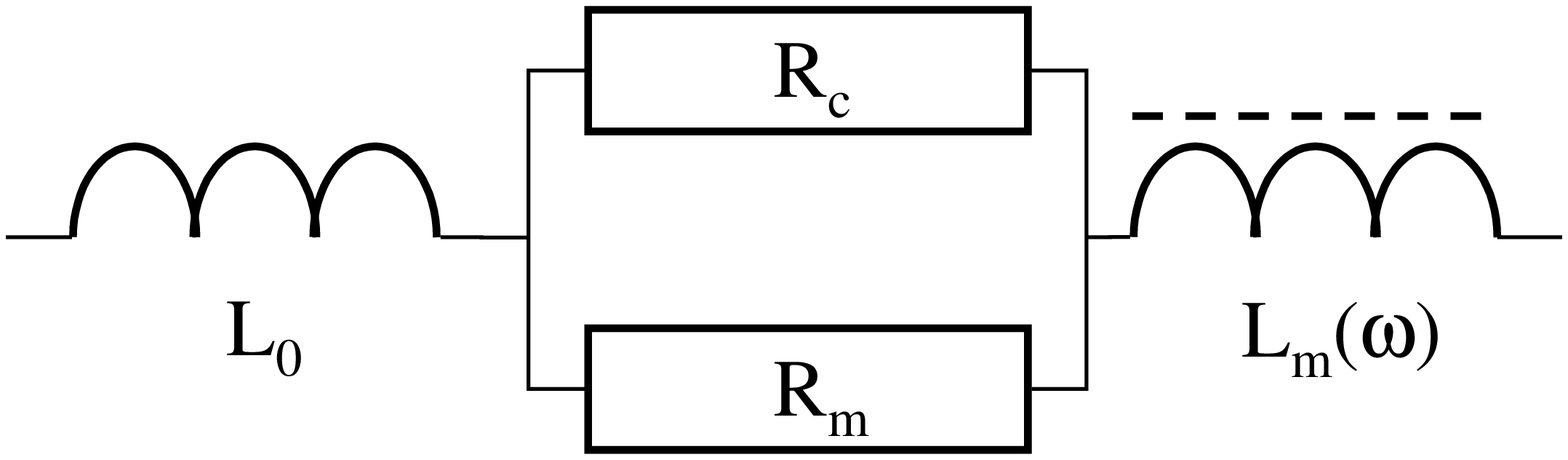}

\caption{-}
\end{figure}


\newpage

\begin{thebibliography}{10}
\bibitem{yam1}M. Yamaguchi, K. Suezawa, Y. Takahashi, K. I. Arai, S. Kikuchi, Y.
Shimada, S. Tanabe, K. Ito, \emph{J. Magn. Magn. Mater.}, \textbf{215-216},
807, 2000. 
\bibitem{review}V. Korenivski, J. Magn. Magn. Mater. \textbf{215-216} (2000) 800. 
\bibitem{saito}T. Saito, K. Tsutsui, S. Yahagi, Y. Matsukura, H. Endoh, T. Eshita,
and K. Hikosaka, \emph{IEEE Trans. Magn.}, \textbf{35}(5), 3187, 1999. 
\bibitem{vk}V. Korenivski and R. B. van Dover, \emph{J. Appl. Phys.}, \textbf{82}(10),
5247, 1997. 
\bibitem{ozawa}K. Ozawa, T. Yanada, H. Masuya, M. Sato, S. Ishio, and M. Takahashi,
\emph{J. Magn. Magn. Mater.}, \textbf{35}, 289, 1983. 
\bibitem{jo}S. Jo, Y. Choi, and S. Ryu, \emph{IEEE Trans. Magn.}, \textbf{33}(5)\textbf{,}
3634, 1997. 
\bibitem{tim}T. J. Klemmer, K. A. Ellis, L. H. Chen, R. B. van Dover, and S. Jin,
\emph{J. Appl. Phys.}, \textbf{87}(2), 830 , 2000. 
\bibitem{McMichael}R. D. McMichael, C. G. Lee, J. E. Bonevich, P. J. Chen, W. Miller,
and W. F. Egelhoff, Jr, J. Appl. Phys. \textbf{89} (2000) 5296. 
\bibitem{grom}A. Gromov, V. Korenivski, D.Haviland, R. B. van Dover, \emph{J. Appl.
Phys.}, \textbf{85}(8), 5202, 1999. 
\bibitem{sukstan}A. Sukstanskii, V. Korenivski, and A. Gromov, J. Appl. Phys. \textbf{89}
(2001) 775. 
\bibitem{Ldc}Inductance calculations: working formulas and tables, Frederick W.
Grover, New York: Van Nostrand, 1946. 
\bibitem{riet}E. van de Riet and F. Roozeboom, \emph{J. Appl. Phys.}, \textbf{81}(1),
350, 1997. 
\bibitem{usov}N. A. Usov, A. S. Antonov and A. N. Lagarkov, \emph{J. Magn. Magn.
Mater.}, \textbf{185,} 159, 1998. 
\end{thebibliography}
\end{document}